\documentclass[11pt]{article}
\usepackage{hyperref}
\pdfoutput=1
\begin{document}
\title{High-frequency capillary waves \\ excited by oscillating microbubbles}
\author{A. Pommella, J. Lantz, V. Poulichet, V. Garbin \\
\\\vspace{6pt} Department of Chemical Engineering, \\ Imperial College London, London SW7 2AZ, UK}
\maketitle
\begin{abstract}
\noindent This fluid dynamics video shows high-frequency capillary waves excited by the volumetric oscillations of microbubbles near a free surface. The frequency of the capillary waves is controlled by the oscillation frequency of the microbubbles, which are driven by an ultrasound field at 37 kHz. Bubble dynamics and capillary wave propagation are imaged by high-speed video microscopy at 300,000 frames per second. Radial capillary waves produced by single bubbles and interference patterns generated by the superposition of waves from multiple bubbles are shown. This video is an entry for the 2013 Gallery of Fluid Motion.
\end{abstract}
% main text
\section*{Introduction}
Ultrasound fields can excite high-frequency capillary waves \cite{Attinger, Yeo}. This fluid dynamics video shows high-frequency capillary waves excited by a microbubble oscillating just beneath the free surface of the fluid. Volumetric oscillations of the bubble are driven by an ultrasound field. The fluid displaced during expansion and compression of the bubble causes a periodic interface perturbation, and the propagation of capillary waves. The frequency of the capillary waves is controlled by the oscillation frequency of the microbubbles.
\section*{Experiments}
Air bubbles (100-250~$\mu$m) are formed in the fluid and allowed to rise to the free surface due to buoyancy. We use a moderately viscous liquid (10-100~mPa~s) to slow down the drainage of the intervening liquid film and prevent the bubbles from immediately bursting. The Newtonian fluids used are water-glycerol mixtures, and water with a viscosity thickener (Laponite S482, Rockwood). The density $\rho$ is in the range 1000-1100~kg~m$^{-3}$ and the surface tension $\gamma$ of the liquid-air interface is in the range 60-70$\times 10^{-3}$~N/m. Ultrasound waves at 37 kHz are transmitted to the sample by a piezoelectric disc driven by a waveform generator and a power amplifier. Visualization is performed in the plane perpendicular to the free interface. The spatial resolution is 2~$\mu$m at 10$\times$ magnification. The bubble dynamics and the propagation of capillary waves are recorded using a high-speed camera at 300,000 frames per second.
\section*{Videos}
Radial capillary waves produced by single bubbles and interference patterns generated by the superposition of waves from multiple bubbles are shown.  The observed wavelength of the waves is $\approx 80\ \mu$m, to be compared with the value predicted from the dispersion relation for capillary waves \cite{Landau} for $f=37$ kHz, $\lambda = \left(\frac{2\pi\gamma}{\rho f^2}\right)^{\frac{1}{3}}\approx 100\ \mu\mathrm{m}$. The observed speed of propagation is $\approx 3$ m/s. A high-resolution and a low-resolution video (\href{./anc/V102379-HighFrequencyCapillaryWaves-Large.mpg}{Video 1} and \href{./anc/V102379-HighFrequencyCapillaryWaves-Small.mpg}{Video 2}) are included. 

%\section{Results}
\section*{Acknowledgements}
The authors acknowledge partial support from EU/FP7 Grant no. 618333.
\bibliographystyle{alpha}

\end{document}